\documentstyle[11pt,newpasp,twoside,epsf]{article}
\reversemarginpar

\begin{document}
\title{The Parkes Multibeam Pulsar Survey Data Release} 
\author{J. Bell$^1$, R. Manchester$^1$, F. Crawford$^3$, A. Lyne$^2$,
F. Camilo$^2$, V. Kaspi$^3$, I. Stairs$^2$, D. Morris$^2$, N. D'Amico$^4$,
N. McKay$^2$, M. Kramer$^2$, D. Sheppard$^2$, A. Possenti$^4$}
\affil{$^1$Australia Telescope National Facility, CSIRO, PO Box 76, Epping NSW
1710, Australia; ~~jbell@atnf.csiro.au}
\affil{$^2$University of Manchester, Jodrell Bank Observatory, Macclesfield,
Cheshire SK11~9DL, UK}
\affil{$^3$Center for Space Research, MIT, Cambridge MA 02139, USA}
\affil{$^4$Osservatorio Astronomico di Bologna, 40127 Bologna, Italy}

\begin{abstract}
The Parkes multibeam pulsar survey began in 1997 and is now about 50\%
complete. It has discovered more than 400 new pulsars so far, including a
number of young, high magnetic field, and relativistic binary pulsars.
Early results, descriptions of the survey and follow up timing programs can
be found in papers by Lyne et al. (1999 MNRAS in press, astro-ph/9911313), Camilo et al. (this
volume, astro-ph/9911185), and Manchester et al. (this volume, astro-ph/9911319). This paper describes the data
release policy and how you can gain access to the raw data and details on
the pulsars discovered.
\end{abstract}

\section{Conditions of Release}

Details of pulsars discovered in the survey are placed on the WWW at the
time of acceptance of the paper announcing the discoveries, or 18 months
after confirmation of the detection, whichever is first. Raw data tapes from
the survey are made available for copying two years after recording.
Details of pulsars observations which fall into these categories can be
found on http://www.atnf.csiro.au/ $\sim$pulsar/psr/pmsurv/pmwww/.  There is
no observatory-based archive, so access to the data should be negotiated
with project PIs.

\section{Volume of data}

One-bit samples are recorded at a rate of 4 kHz for each of the 96 channels
per beam.  Each 35 minute survey observation with the 13-beam system fills
1.3 GBytes ($\sim$100 MBytes per beam) in its raw form. This is recorded on
DLT7000 tapes which hold up to 35 GBytes and cost around US\$80 each. About
3000 such observations will be made in completing the survey, giving a total
data set of about 4 TBytes. Our search processing requires 130 hours of CPU
time on a SUN Ultra 1 for each 35 minute observation.

\section{How to Access data}

Data logs are available on {\small
http://www.atnf.csiro.au/$\sim$pulsar/psr/pmsurv/pmwww/.} For the multibeam
pulsar survey and followup timing projects the PIs to contact are Andrew
Lyne (agl@jb.man.ac.uk) or Dick Manchester (rmanches@atnf.csiro.au) or
Fernando Camilo (fernando@astro.columbia.edu).

\begin{description}

\item{\bf Small requests:} If you want a copy of a single tape or a few
individual observations we are happy to copy the data and post you a tape or
a CD. Please indicate your preference for media type. 

\item{\bf Larger requests:} You would need to come to Epping or Jodrell Bank
and copy the data onto your own DLT7000 media. If you want a copy of the
complete survey, you would need to bring a workstation, a DLT7000 tape drive
and a drive for the media type of your choice.

\item{\bf Timing Data:} Processed archives or TOAs can be made available by
ftp, CD or exabyte. For raw data requests the above schemes for small and
large requests would apply.

\end{description}

\section{Has the position you are interested in been observed ?}

The observing logs on the web site contain grid IDs for the centre beam of
each 35 minute survey observation. The grid ID $= il*1000+ib$ where
$b=(ib-500)*0.20207$ and $l=(il-5000+0.5*{\rm mod}(ib,2))*0.2333$. The
Galactic centre has grid ID 5000500. The nominal centre positions of the
other 12 beams can be determined from the offsets relative to beam 1 given
in the table below.

\footnotesize{
\noindent
\begin{tabular}{lrrrrrrrrrrrrr}\hline
     &     2 &     3 &     4 &     5 &     6 &     7 &     8 &
9 &     10 &     11 &    12 &    13 \\ \hline 
l    & -0.24 &  0.24 &  0.49 &  0.24 & -0.24 & -0.49 & -0.73 &
0.00 &  0.73 &  0.73 &  0.00 & -0.73 \\
b    &  0.42 &  0.42 &  0.00 & -0.42 & -0.42 &  0.00 &  0.42 &  
0.85 &  0.42 & -0.42 & -0.85 & -0.42 \\ \hline
\end{tabular}
}

\normalsize
\section{Software Tools}

We guarantee to provide software to read the tapes on a SUN Ultra class
workstation. This software also works on most brands of UNIX operating
systems. A range of software tools are available on an all care and no
responsibility basis.

\begin{tabular}{ll}
pmfind	& search for pulsars \\
pdm	& fold and dedisperse data for a given pulsar \\
fch3	& fold data with precision suitable for timing \\
tarch   & form processed archives for timing analysis \\
treduce	& analyse timing archives \\
pmhex	& survey observation database \\
foldch	& fold and analyse individual filter bank channels \\
\end{tabular}

\section{Collaborative Projects}

It is possible to get access to data and results from the survey earlier by
collaborative arrangements. To date we have embarked on such arrangements
with 11 groups and welcome proposals from others.

\end{document}